\documentclass[twocolumn,showpacs,preprintnumbers,amsmath,amssymb]{revtex4}
\usepackage{amssymb}
\usepackage[dvips]{graphicx}
\begin{document}
\title{Superlight bipolarons and a checkerboard d-wave condensate in cuprates}
\author{A.S.~Alexandrov}
\address{Department of Physics, Loughborough University,Loughborough LE11\,3TU, UK}

\begin{abstract}
%\large
The seminal work by Bardeen, Cooper and Schrieffer taken further by
Eliashberg to the intermediate coupling solved the problem of conventional
superconductors about half a century ago. The Fr\"{o}hlich and Jahn-Teller
electron-phonon interactions were identified as an essential piece of
physics in all novel superconductors. The BCS theory provides a
qualitatively correct description of some of them like magnesium diborade
and doped fullerenes (if the polaron formation is taken into account).
However, cuprates remain a problem. Here I show that the bipolaron extension
of the BCS theory to the strong-coupling regime could be a solution.
Low-energy physics in this regime is that of small 'superlight' bipolarons,
which are real-space mobile bosonic pairs dressed by phonons. The symmetry
and space modulations of the order parameter are explained in the framework
of the bipolaron theory. A d-wave Bose-Einstein condensate of bipolarons
reveals itself as a checkerboard modulation of the hole density and of the
gap below Tc.
\end{abstract}

\vskip2pc]

\maketitle
%\Large
%\narrowtext

\bigskip

\section{Introduction: pairing is individual in many cuprates}

Experimental \cite{mih,cal,tim,guo,shen2,ega,mul2,she2} and theoretical \cite
{phil,alemot,allen3,gor4,dev2} evidence for an exceptionally strong
electron-phonon interaction in high temperature superconductors is now so
overwhelming, that even some advocates of nonphononic mechanisms \cite{kiv}
accept this fact. Our view is that the extension of the BCS theory towards
the strong interaction between electrons and ion vibrations describes the
phenomenon naturally. High temperature superconductivity exists in the
crossover region of the electron-phonon interaction strength from the
BCS-like to bipolaronic superconductivity as was predicted before \cite{ale0}%
, and explored in greater detail after the discovery \cite
{alemot,alej,rice,emi0,dev,workshop,micr}.

Polarons form the Cooper pairs, if their coupling is not strong enough.
However, real-space small bipolarons are also feasible due to a polaron
collapse of the Fermi energy. There is strong evidence for the Bose-Einstein
condensation of small bipolarons in many cuprates, in particular at low
doping. Let us, for example, estimate a renormalised Fermi energy. The band
structure is quasi-two-dimensional with a few degenerate hole pockets in
cuprates. Applying the parabolic approximation for the band dispersion we
obtain the {\it renormalized} Fermi energy as 
\begin{equation}
\epsilon _{F}={\frac{\pi n_{i}d}{{m_{i}^{\ast }}}},
\end{equation}
where $d$ is the interplane distance, and $n_{i},m_{i}^{\ast }$ are the
density of holes and their effective mass in each of the hole subbands $i$
renormalized by the electron-phonon (and by any other) interaction. One can
express the renormalized band-structure parameters through the in-plane
magnetic-field penetration depth at $T\approx 0$, measured experimentally: 
\begin{equation}
\lambda _{H}^{-2}=4\pi e^{2}\sum_{i}\frac{n_{i}}{m_{i}^{\ast }}.
\end{equation}
As a result, we obtain a {\em parameter-free }expression for the ``true''
Fermi energy as 
\begin{equation}
\epsilon _{F}={\frac{d}{{4ge^{2}\lambda _{H}^{2}}}},
\end{equation}
where $g$ is the degeneracy of the spectrum, which may depend on doping. One
expects $4$ hole pockets inside the Brillouin zone (BZ) due to the
Mott-Hubbard gap in underdoped cuprates. If the hole band minima are shifted
with doping to BZ boundaries, all their wave vectors would belong to the
stars with two or more prongs. The groups of wave vectors for these stars
have only 1D representations. It means that the spectrum will be degenerate
with respect to the number of prongs, i.e $g\geqslant 2$. Because Eq.(3)
does not contain any other band-structure parameters, the estimate of $%
\epsilon _{F}$ using this equation does not depend very much on the
parabolic approximation for the band dispersion.

Generally, the ratios $n/m^{\ast }$ in Eq.(1) and Eq.(2) are not necessarily
the same. The `superfluid' density in Eq.(2) might be different from the
total density of delocalized carriers in Eq.(1). However, in a
translationally invariant system they must be the same \cite{pop}. This is
also true even in the extreme case of a pure two-dimensional superfluid,
where quantum fluctuations are important. One can, however, obtain a reduced
value of the zero temperature superfluid density in the dirty limit, $l\ll
\xi (0)$, where $\xi (0)$ is the zero-temperature coherence length. The
latter was measured directly in cuprates as the size of the vortex core. It
is about 10 $\AA $ or even less. On the contrary, the mean free path was
found surprisingly large at low temperatures, $l\sim $ 100-1000 $\AA $.
Hence, it is rather probable that novel superconductors are in the clean
limit, $l\gg \xi (0)$, so that the parameter-free expression for $\epsilon
_{F}$, Eq.(3), is perfectly applicable.

A parameter-free estimate of the Fermi energy obtained using Eq.(3) is
presented in the Table. 
\begin{table}[tbp]
\caption{The Fermi energy (multiplied by the degeneracy) of cuprates}
\begin{tabular}[t]{llllllll}
Compound & $T_{c}$ (K) & $\lambda _{H,ab}$ $(\AA )$ & d$(\AA )$ & $%
g\epsilon_{F}$ (meV) &  &  &  \\ \hline
$La_{1.8}Sr_{0.2}CuO_{4}$ & 36.2 & 2000 & 6.6 & 112 &  &  &  \\ 
$La_{1.78}Sr_{0.22}CuO_{4}$ & 27.5 & 1980 & 6.6 & 114 &  &  &  \\ 
$La_{1.76}Sr_{0.24}CuO_{4}$ & 20.0 & 2050 & 6.6 & 106 &  &  &  \\ 
$La_{1.85}Sr_{0.15}CuO_{4}$ & 37.0 & 2400 & 6.6 & 77 &  &  &  \\ 
$La_{1.9}Sr_{0.1}CuO_{4}$ & 30.0 & 3200 & 6.6 & 44 &  &  &  \\ 
$La_{1.75}Sr_{0.25}CuO_{4}$ & 24.0 & 2800 & 6.6 & 57 &  &  &  \\ 
$YBa_{2}Cu_{3}O_{7}$ & 92.5 & 1400 & 4.29 & 148 &  &  &  \\ 
$YBaCuO(2\%Zn)$ & 68.2 & 2600 & 4.29 & 43 &  &  &  \\ 
$YBaCuO(3\%Zn)$ & 55.0 & 3000 & 4.29 & 32 &  &  &  \\ 
$YBaCuO(5\%Zn)$ & 46.4 & 3700 & 4.29 & 21 &  &  &  \\ 
$YBa_{2}Cu_{3}O_{6.7}$ & 66.0 & 2100 & 4.29 & 66 &  &  &  \\ 
$YBa_{2}Cu_{3}O_{6.57}$ & 56.0 & 2900 & 4.29 & 34 &  &  &  \\ 
$YBa_{2}Cu_{3}O_{6.92}$ & 91.5 & 1861 & 4.29 & 84 &  &  &  \\ 
$YBa_{2}Cu_{3}O_{6.88}$ & 87.9 & 1864 & 4.29 & 84 &  &  &  \\ 
$YBa_{2}Cu_{3}O_{6.84}$ & 83.7 & 1771 & 4.29 & 92 &  &  &  \\ 
$YBa_{2}Cu_{3}O_{6.79}$ & 73.4 & 2156 & 4.29 & 62 &  &  &  \\ 
$YBa_{2}Cu_{3}O_{6.77}$ & 67.9 & 2150 & 4.29 & 63 &  &  &  \\ 
$YBa_{2}Cu_{3}O_{6.74}$ & 63.8 & 2022 & 4.29 & 71 &  &  &  \\ 
$YBa_{2}Cu_{3}O_{6.7}$ & 60.0 & 2096 & 4.29 & 66 &  &  &  \\ 
$YBa_{2}Cu_{3}O_{6.65}$ & 58.0 & 2035 & 4.29 & 70 &  &  &  \\ 
$YBa_{2}Cu_{3}O_{6.6}$ & 56.0 & 2285 & 4.29 & 56 &  &  &  \\ 
$HgBa_{2}CuO_{4.049}$ & 70.0 & 2160 & 9.5 & 138 &  &  &  \\ 
$HgBa_{2}CuO_{4.055}$ & 78.2 & 1610 & 9.5 & 248 &  &  &  \\ 
$HgBa_{2}CuO_{4.055}$ & 78.5 & 2000 & 9.5 & 161 &  &  &  \\ 
$HgBa_{2}CuO_{4.066}$ & 88.5 & 1530 & 9.5 & 274 &  &  &  \\ 
$HgBa_{2}CuO_{4.096}$ & 95.6 & 1450 & 9.5 & 305 &  &  &  \\ \hline
\end{tabular}
\end{table}
The renormalised Fermi energy of about 30 cuprates is less than $100$ $meV$,
in particular if the degeneracy $g\geq 2$ is taken into account. That should
be compared with the characteristic phonon frequency, which is estimated as
the plasma frequency of oxygen ions, 
\begin{equation}
\omega _{0}=\sqrt{\frac{4\pi Z^{2}e^{2}N}{M}}.
\end{equation}
One obtains $\omega _{0}$=$84meV$ with $Z=2$, $N=6/V_{cell}$, $M=16$ $a.u.$
for $YBa_{2}Cu_{3}O_{6}$. Here $V_{cell}$ is the volume of the (chemical)
unit cell. The estimate agrees with the measured phonon spectra. As
established experimentally in cuprates, the high-frequency phonons are
strongly coupled with carriers. Therefore the low Fermi energy is a serious
problem for the BCS (or Migdal-Eliashberg) approach. Since the Fermi energy
is small and phonon frequencies are high, the Coulomb pseudopotential $\mu
_{c}^{\ast }$ is of the order of the bare Coulomb repulsion, $\mu _{c}^{\ast
}\simeq \mu _{c}\simeq 1$ because the Tolmachev logarithm is ineffective.
Hence, to get an experimental T$_{c}$, one has to have a strong coupling, $%
\lambda >\mu _{c}$. However, one cannot increase $\lambda $ without
accounting for the polaron collapse of the band. Even in the region of the
applicability of the Eliashberg theory (i.e. at $\lambda \leq 0.5$), the
non-crossing diagrams cannot be treated as vertex $corrections$, since they
are comparable to the standard terms, if $\omega _{0}/\epsilon _{F}\gtrsim 1$%
. Because novel superconductors are in the nonadiabatic regime, interaction
with phonons must be treated in the framework of the multi-polaron theory at
any value of $\lambda$.

In many cases (Table) the renormalized Fermi energy is so small that pairing
is certainly individual, i.e. the bipolaron radius is smaller than the
inter-carrier distance. Indeed, this is the case, if 
\begin{equation}
\epsilon _{F}\lesssim \pi \Delta .
\end{equation}
The bipolaron binding energy $\Delta $ is thought to be {\it twice} the
so-called pseudogap experimentally measured in the normal state of many
cuprates, $\Delta \gtrsim 100meV,$ so that Eq.(5) is well satisfied in
underdoped and even in a few optimally and overdoped cuprates. One should
notice that the coherence length in the charged Bose gas has nothing to do
with the size of the boson. It depends on the interparticle distance and the
mean-free path, \cite{alemot}, and might be as large as in the BCS
superconductor. Hence, it is incorrect to apply the ratio of the coherence
length to the inter-carrier distance as a criterium of the BCS-Bose liquid
crossover. The correct criterium is given by Eq.(5).

\section{Symmetry of the order parameter}

The anomalous Bogoliubov average 
\[
F_{ss^{\prime }}({\bf r}_{1},{\bf r}_{2})=\left\langle \left\langle \Psi
_{s}({\bf r}_{1})\Psi _{s^{\prime }}({\bf r}_{2})\right\rangle \right\rangle
, 
\]
is the superconducting order parameter both in the weak and strong-coupling
regimes ( $\Psi _{s}({\bf r})$ annihilates a carrier with spin $s$ and
coordinate ${\bf r}$ ). $F_{ss^{\prime }}({\bf r}_{1},{\bf r}_{2})$ depends
on the relative coordinate ${\bf \rho =r}_{1}-{\bf r}_{2}$ of two electrons
of the pair, and on the center-of-mass coordinate ${\bf R}=({\bf r}_{1}+{\bf %
r}_{2})/2$. Hence, its Fourier transform, $f({\bf k,K})$, depends on the
relative momentum ${\bf k}$ and on the center-of-mass momentum ${\bf K.}$ In
the BCS theory, where ${\bf K=}0$ (in a homogeneous superconductor), the
Fourier transform of the order parameter is proportional to the gap in the
quasiparticle excitation spectrum, $f({\bf k,K})\varpropto \Delta _{{\bf k}}$%
. Hence the symmetry of the order parameter and the symmetry of the gap are
the same in the weak-coupling regime. Under the rotation of the coordinate
system, $\Delta _{{\bf k}}$ changes its sign, if the Cooper pairing is
d-wave. In this case the BCS quasiparticle spectrum is gapless.

In the bipolaron theory the symmetry of the Bose-Einstein condensate is not
necessarily the same as the ``internal'' symmetry of a pair \cite{alesym}.
While the latter describes transformation of $f({\bf k,K})$ with respect to
the rotation of ${\bf k}$, the former (``external'') symmetry is related to
the rotation of ${\bf K.}$ Therefore it depends on the bipolaron band
dispersion, but not on the symmetry of the bound state. \bigskip

\begin{figure}
\begin{center}
\includegraphics[angle=-0,width=0.47\textwidth]{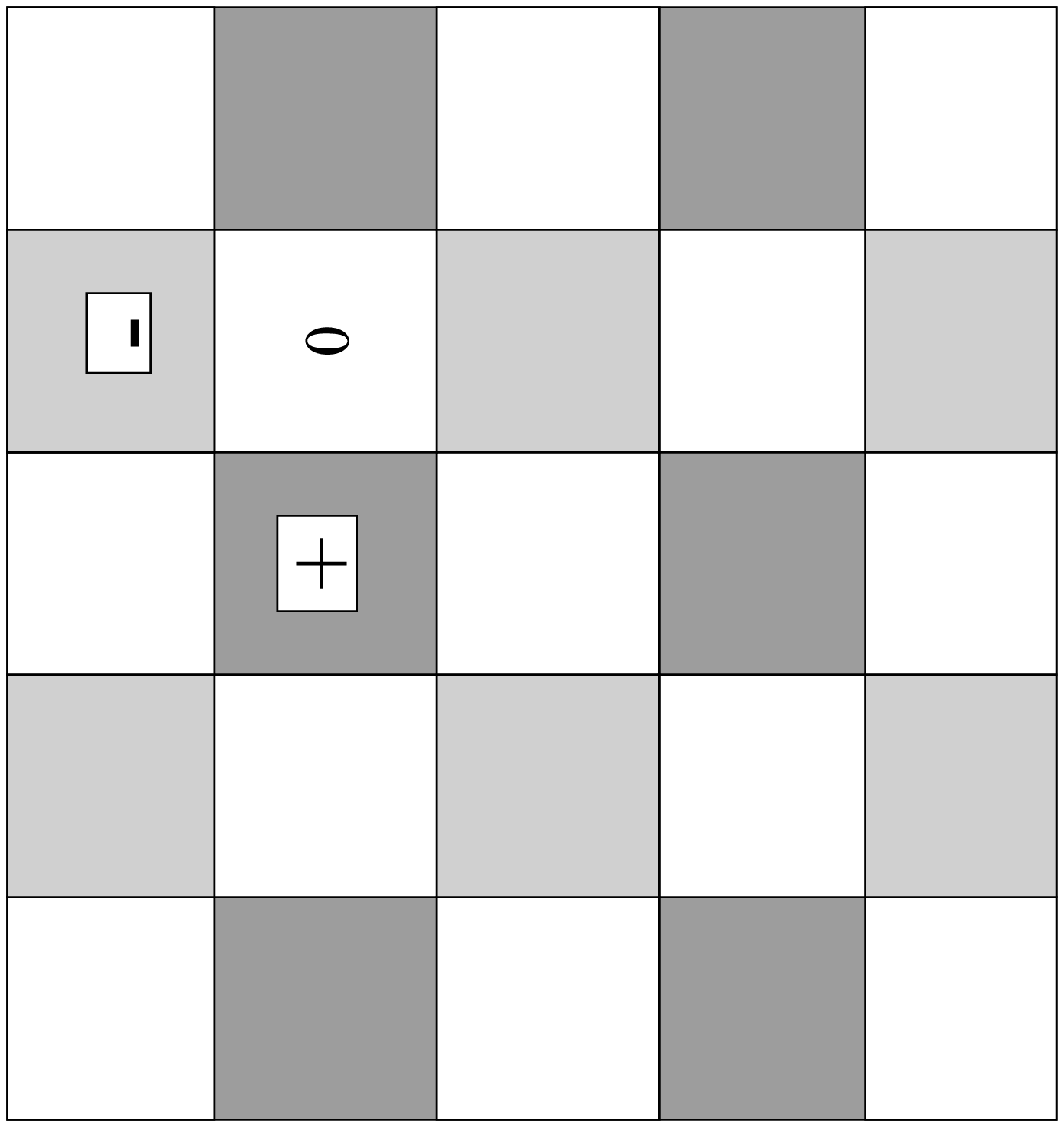}
\vskip -0.5mm
\caption{d-wave condensate wave function, in the Wannier
representation. The order parameter has different signs in the shaded cells
and is zero in the blank cells.}                                                \end{center}
\end{figure}

As an example, let us consider a tight-binding bipolaron spectrum comprising
two bands on a square lattice with the period $a=1,$ 
\begin{eqnarray}
E_{{\bf K}}^{x} &=&t\cos (K_{x})-t^{\prime }\cos (K_{y}), \\
E_{{\bf K}}^{y} &=&-t^{\prime }\cos (K_{x})+t\cos (K_{y}).  \nonumber
\end{eqnarray}
They transform into one another under $\pi /2$ rotation. If $t,t^{\prime
}>0, $ ``$x"$ bipolaron band has its minima at ${\bf K=}(\pm \pi ,0)$ and $y$%
-band at ${\bf K}=(0,\pm \pi )$. These four states are degenerate, so that
the condensate wave function $\psi _{s}({\bf m})$ in the site space, ${\bf m}%
=(m_{x},m_{y}),$ is given by 
\begin{equation}
\psi _{s}({\bf m})=N^{-1/2}\sum_{{\bf K}=(\pm \pi ,0),(0,\pm \pi )}b_{{\bf K}%
}e^{-i{\bf K\cdot m}}.
\end{equation}
where $b_{{\bf K}}=\pm \sqrt{n_{s}}$ are $c$-numbers at $T=0.$ The
superposition, Eq.(7), respects the time-reversal and parity symmetries, if 
\begin{equation}
\psi _{s}^{\pm }({\bf m})=\sqrt{n_{s}}\left[ \cos (\pi m_{x})\pm \cos (\pi
m_{y})\right] .
\end{equation}

Two order parameters, Eq.(8), are physically identical because they are
related by the translation transformation, $\psi _{s}^{+}(m_{x},m_{y})=\psi
_{s}^{-}(m_{x},m_{y}+1)$. Both have a $d$-wave symmetry changing sign, when
the lattice is rotated by $\pi /2$, Fig.1. The $d$-wave symmetry is entirely
due to the bipolaron energy dispersion with four minima at ${\bf K\neq 0}$.
When the bipolaron spectrum is not degenerate and its minimum is located at $%
\Gamma $ point of the Brillouin zone, the condensate wave function is $s$%
-wave with respect to the center-of-mass coordinate. The symmetry of the
normal state (pseudo)gap has little to do with the symmetry of the order
parameter in the strong-coupling regime. The one-particle pseudogap is half
of the bipolaron binding energy $\Delta /2$, and does not depend on any
momentum in zero order of the polaron bandwidth, i.e. it has an ``$s"$-wave
symmetry. In fact, due to a finite dispersion of polaron and bipolaron
bands, the pseudogap is an $anysotropic$ s-wave. A multi-band electron
structure can include bands weakly coupled with phonons which could overlap
with the bipolaronic band. In this case CBG coexists with the Fermi gas,
like $^{4}He$ bosons coexist with $^{3}He$ fermions in the mixture of
Helium-4 and Helium-3 \cite{alemot}. The normal state one-particle
excitation spectrum of such mixtures is gapless.

\section{Checkerboard modulations of the superconducting gap}

Independent observations of normal state pseudogaps in a number of magnetic
and kinetic measurements, and unusual critical phenomena tell us that many
cuprates may not be $BCS$ superconductors. Indeed their superconducting
state is as anomalous as the normal one. In particular, there is strong
evidence for a $d$-like order parameter (changing sign when the $CuO_{2}$
plane is rotated by $\pi /2$) in cuprates \cite{ann}. A number of
phase-sensitive experiments \cite{pha} provide unambiguous evidence in this
direction; furthermore, the low temperature magnetic penetration depth \cite
{bonn,xia} was found to be linear in a few cuprates as expected for a d-wave
BCS superconductor. However, different tunnelling spectroscopies, in
particular $c$-axis Josefson tunnelling \cite{klem}, and some high-precision
magnetic measurements \cite{mulsym} show a more usual $s$-like symmetry.
Other studies even reveal an upturn in the temperature dependence of the
penetration depth below some characteristic temperature \cite{wal}. Also
both angle-resolved photoemission (ARPES) and tunnelling spectroscopies
often show a very large energy gap with $2\Delta /T_{c}$ ratio well above
that expected in any-coupling BCS theory.

Strong deviations from the Fermi/BCS-liquid behaviour are suggestive of a
new electronic state in cuprates, which could be a charged Bose liquid of
bipolarons. In the bipolaron theory the symmetry of the Bose-Einstein
condensate on a lattice should be distinguished from the `internal' symmetry
of a single-bipolaron wave function, and from the symmetry of a
single-particle excitation gap. As described above the Bose-Eistein
condensate of bipolarons could be $d$-wave, if bipolaron bands have their
minima at finite ${\bf K}$ in the center-of-mass Brillouin zone. At the same
time the single-particle excitation spectrum is an anysotropic $s$-wave
providing a possible explanation of conflicting experimental observations.

The two-dimensional patterns, Fig.1, are oriented along the diagonals, i.e.
the $d$-wave bipolaron condensate is `striped'. Hence there is a fundamental
connection between stripes detected by different techniques \cite{tran,bia}
in cuprates and the symmetry of the order parameter \cite{alesym}.
Originally antiferromagnetic interactions were thought to give rise to spin
and charge segregation (stripes) \cite{zaa}. However the role of long-range
Coulomb and Fr\"{o}hlich interactions was not properly addressed. We showed
that the Fr\"{o}hlich electron-phonon interaction combined with the direct
Coulomb repulsion does not lead to charge segregation like strings or
stripes in doped insulators, and the antiferromagnetic exchange interaction
is not sufficient to produce long stripes either \cite{alekabstr}.

If (bi)polaronic carriers in many cuprates are in a liquid state, one can
pose a key question of how one can see stripes at all. Actually, the
bipolaron condensate is striped owing to the bipolaron energy band
dispersion, as discussed above. In this scenario the hole density, which is
about twice of the condensate density at low temperatures, is striped, with
the characteristic period of stripes determined by inverse wave vectors
corresponding to bipolaron band-minima. Such interpretation of stripes is
consistent with the inelastic neutron scattering in $YBa_{2}Cu_{3}O_{7-%
\delta }$ , where the incommensurate peaks were observed $only$ in the $%
superconducting$ state \cite{bou}. The vanishing at $T_{c}$ of the
incommensurate peaks is inconsistent with any other stripe picture, where a
characteristic distance needs to be observed in the normal state as well. On
the contrary, with the d-wave $striped$ Bose-Einstein condensate the
incommensurate neutron peaks should disappear above $T_{c}$, as observed.
Also recent X-ray spectroscopy \cite{boz} did not find any bulk charge
segregation in the {\it normal} state of a high-$Tc$ $LaCuO_{4+\delta}$
suggesting an absence of in-plane carrier density modulations in this
material above $T_{c}$.

The $d$-wave condensate reveals itself as a {\it checkerboard} real-space
modulation of the hole density, Fig.1. Then the superconducting gap should
be modulated as well. Indeed, as shown in Ref.\cite{aleand}, the
single-particle excitation spectrum of the bipolaronic superconductor can be
obtained using the BCS excitation spectrum but with a negative chemical
potential, 
\begin{equation}
\epsilon ({\bf k})=\left[ (k^{2}/2m^{\ast }+\Delta _{p})^{2}+\Delta
_{c}(T)^{2}\right] ^{1/2},
\end{equation}
where $\Delta _{c}(T)^{2}=constant\times |\psi _{s}|^{2}$ is a coherent
component of the gap. This spectrum is quite different from the BCS
quasiparticles because the chemical potential is negative with respect to
the bottom of the single-particle band, $\mu =-\Delta _{p}$. A single
particle gap, $\Delta /2$, defined as the minimum of $\epsilon ({\bf k})$,
is given by 
\begin{equation}
\Delta /2=\left[ \Delta _{p}^{2}+\Delta _{c}(T)^{2}\right] ^{1/2}.
\end{equation}
It varies with temperature from $\Delta (0)/2=\left[ \Delta _{p}^{2}+\Delta
_{c}(0)^{2}\right] ^{1/2}$ at zero temperature down to the temperature
independent $\Delta _{p}$ above $T_{c}$. The spectrum, Eq.(10) was obtained
for a homogeneous condensate. However, in the framework of a semiclassical
approximation, we can apply Eq.(10) also with a modulated $\psi _{s}({\bf m})
$, if the period of modulations is large enough \cite{alekab}. As a result,
the superconducting gap is modulated. Importantly, the checkerboard
modulation of the gap have been observed in the tunnelling experiments with 
\cite{hoff} and without \cite{kapit,hoff2} applied magnetic field in $Bi$
cuprates.

In conclusion, I have shown  that there is a real-space pairing in many
cuprates due to the polaron collapse of the Fermi energy. Remarkably, the
bipolaron theory provides an explanation of the d-wave order parameter, the
anysotropic s-wave single-particle gap, and its stripe checkerboard
modulations below $T_{c}$ within a single microscopic approach.

This work has been supported by the Leverhulme Trust (grant F/00261/H). I
greatly appreciate enlightening discussions with I. Bozovic, \ K. McElroy,
V. V. Kabanov, \ A. J. Leggett, D. Mihailovic, and K. A. M\"{u}ller.

\end{document}